\documentclass[aps,pra,superscriptaddress,twocolumn,showpacs]{revtex4}
\usepackage[usenames,dvipsnames]{color}

\usepackage{epsfig,epstopdf}
\usepackage{bm,epstopdf}
\usepackage{graphicx,times}
\usepackage[usenames]{color}
\usepackage{amssymb,amsmath,bbm,amsfonts,amsthm}
\usepackage{subfigure}

\newcommand{\ket}[1]{\left\vert#1\right\rangle}

\newcommand{\Sprod}[2]{\langle#1\vert#2\rangle}
\newcommand{\SprodO}[3]{\langle#1\vert#2\vert#3\rangle}

\newcommand{\bra}[1]{\left\langle#1\right\vert}

\newcommand{\valmed}[1]{\left\langle#1\right\rangle}
\DeclareMathOperator{\Tr}{Tr}

\begin{document}

\title {Effective cutting of a quantum spin chain by bond impurities}

\author{T.~J.~G. Apollaro}
\affiliation{Dipartimento di Fisica \& INFN--Gruppo collegato di Cosenza, Universit\`a della Calabria,
       Via P. Bucci, 87036 Arcavacata di Rende (CS), Italy}

\affiliation{Centre for Theoretical Atomic, Molecular, and Optical Physics,
   School of Mathematics and Physics, Queen's University Belfast,
   BT7\,1NN, United Kingdom}

\author{F. Plastina}
\affiliation{Dipartimento di Fisica \& INFN--Gruppo collegato di Cosenza, Universit\`a della Calabria,
       Via P. Bucci, 87036 Arcavacata di Rende (CS), Italy}

\author{L. Banchi}
\affiliation{ Department of Physics and Astronomy, University College
London, Gower St., London WC1E 6BT, United Kingdom }
\affiliation{ISI Foundation, Via Alassio 11/c,
       I-10126 Torino (TO), Italy}

\author{A. Cuccoli}
\affiliation{Dipartimento di Fisica, Universit\`a di Firenze,
       Via G. Sansone 1, I-50019 Sesto Fiorentino (FI), Italy}
\affiliation{INFN Sezione di Firenze, via G.Sansone 1,
       I-50019 Sesto Fiorentino (FI), Italy}

\author{R. Vaia}
\affiliation{Istituto dei Sistemi Complessi,
       Consiglio Nazionale delle Ricerche,
       via Madonna del Piano 10,
       I-50019 Sesto Fiorentino (FI), Italy}
       \affiliation{INFN Sezione di Firenze, via G.Sansone 1,
       I-50019 Sesto Fiorentino (FI), Italy}

\author{P. Verrucchi}
\affiliation{Istituto dei Sistemi Complessi,
       Consiglio Nazionale delle Ricerche,
       via Madonna del Piano 10,
       I-50019 Sesto Fiorentino (FI), Italy}
\affiliation{Dipartimento di Fisica, Universit\`a di Firenze,
       Via G. Sansone 1, I-50019 Sesto Fiorentino (FI), Italy}
\affiliation{INFN Sezione di Firenze, via G.Sansone 1,
       I-50019 Sesto Fiorentino (FI), Italy}

\author{M. Paternostro}
\affiliation{Centre for Theoretical Atomic, Molecular, and Optical Physics,
   School of Mathematics and Physics, Queen's University Belfast,
   BT7\,1NN, United Kingdom}
   \affiliation{Institut f\"ur Theoretische Physik, Albert-Einstein-Allee 11, Universit\"at Ulm, D-89069 Ulm, Germany}


\begin{abstract}

Spin chains are promising media for short-haul quantum communication.
Their usefulness is manifested in all those situations where
stationary information carriers are involved. In the majority of the
communication schemes relying on quantum spin chains, the latter are
assumed to be finite in length, with well addressable end-chain
spins. In this paper we propose that such configuration
could actually be achieved by a mechanism that is able to effectively
cut a spin ring through the
insertion of bond defects. We then show how suitable physical
quantities can be identified as figures of merit for the
effectiveness of the cut. We find that, even for modest strengths of
the bond defect, a ring is effectively cut at the defect site. In
turn, this has important effects on the amount of correlations shared
by the spins across the resulting chain, which we study by means of a
scattering-based mechanism of a clear physical interpretation.
\end{abstract}

\date{\today}

\maketitle

\label{S.Intro}

In the last decade, the idea of connecting stationary information carriers via 
one-dimensional spin systems has been developed significantly and several
strategies have been proposed for obtaining high-quality quantum-state
and entanglement transfer, as well as entangling gates~\cite{refs1,refs2}. The general paradigm
involves two remote qubits located at each end of a chain of interacting
spins mediating the exchange of information between the distant particles.
Together with the strength of the intra-chain coupling, the length of
the chain, as measured for instance by the number of its spins, is a
key parameter that determines the operational time and quality of a
given communication scheme. In fact, in any practical implementation,
the spin-chain medium needs to be of finite length with well
identified and addressable first ({\it head}) and last (\emph{tail})
elements.

Depending on the actual physical realization, one can think of
different ways of fulfilling such requirements. In this paper we
consider the case of a medium modelled by a chain of interacting
spin-$1/2$ particles, such as the crystals listed in Table 1 of
Ref.~\cite{MikeskaS91} or the more recently proposed molecular
rings~\cite{Timco2009, TroianiBCLA2010}. Other physical
realizations, ranging from ultracold-atom systems to
coupled-cavity arrays~\cite{Lewensteinetal07, GiampaoloI2010}
adhere well to such a model. We specifically address the problem
of obtaining a one-dimensional spin system of finite length and
open boundary conditions (OBC), hereafter called ``segment'', out
of a spin chain with periodic boundary conditions (PBC). As the
latter structure can be generally represented as a closed ring (of
either finite or infinite length), we will refer to the above
problem as that of ``cutting a ring''.

As a ring-cutting mechanism basically changes PBC into OBC, and
relying on general arguments about how impurities affect the
behavior of one-dimensional systems, we propose the insertion of
one impurity as an effective tool for realizing one cut. In
particular, we consider the case when the impurity corresponds to
a variation of the interaction strength between two neighboring
spins, with respect to the otherwise homogeneous couplings. The
effect of the presence of this kind of bond-impurity on the ground
state of the antiferromagnetic XXZ Heisenberg spin-$\frac{1}{2}$
model has been investigated via renormalization group techniques
in Refs.~\cite{EggertA1992,RommerE2000,SchusterE2002}, where it
has been shown that this kind of impurity embodies a relevant
perturbation and yields to a fixed point in the
renormalization-group flow corresponding to OBC for an infinite
interaction's strength. In this paper we solve analytically the
impurity XX Heisenberg spin-$\frac{1}{2}$ model via the
Jordan-Wigner mapping into a non-interacting spinless fermionic
model and determine quantitatively the cutting effect for finite
interaction's strength via quantum-information inspired figures of
merit, such as classical and quantum correlations and fidelity
measures.

An equally important motivation to investigate the impurity-driven
ring-cutting mechanism is to analyze the emergence of boundary
effects, such as Friedel-like oscillations~\cite{friedel} of the
fermion density, \textit{i.e.}, the local magnetization, driven by
the impurity strength. These effects are more pronounced in proximity
of the impurity spins~\cite{Baskaran79, Gildenblat1984, ShinkevichSE11}. They allow for the tuning of
the degree of entanglement shared by two arbitrary spins of the medium (even
different from the impurities) along the lines of Refs.~\cite{OsendaHK2003,ApollaroCFPV2008,ApollaroCFPV2008.2,PlastinaA2007}.

The paper is organized as follows: In Sec.~\ref{S.model} we
introduce the specific model addressed here, namely that of a ring
of $2M$ spin-$1/2$ particles, interacting via a nearest-neighbor,
planar and isotropic magnetic exchange model, hereafter referred
to as XX interaction. The effect of an inserted bond impurity is
here analytically studied in the thermodynamic limit
$M{\to}\infty$. In Sec.~\ref{S.cutting-2points} we discuss the
thermodynamic-limit behavior of in- and out-of-plane magnetic
correlations, concurrence~\cite{Wootters98}, and quantum
discord~\cite{discord1,discord2,discordreview}, which are some of
our elected figures of merit for the characterization of the
ring-cutting mechanism. The case of finite $M$ is considered in
Sec.~\ref{S.fidelity} where we study the fidelity~\cite{Josza94}
between the ground state of a ring that has been effectively cut
by a bond impurity, and that of the segment it should mimic. The
overall analysis is carried out as the parameter characterizing
the bond impurity is varied and, as far as the finite-length case
is concerned, for different values of $M$. Finally, we draw our
conclusions in Sec.~\ref{S.conclusions}.

\section{The Model}
\label{S.model}

We consider a one-dimensional system of $2M$ ($M\in\mathbb{N}$)
interacting spin-$1/2$ particles in the presence of a uniform
magnetic field. The interaction is of the isotropic and planar
(XX) Heisenberg form
\begin{equation}
\label{e.hamilton0}
 \hat{\cal H}_0{=}-\frac{J}{2}\sum_{n{=}-M+\frac12}^{M-\frac12}
 \big(\hat \sigma^x_n \hat
 \sigma^x_{n+1} {+}\hat\sigma^y_n\hat\sigma^y_{n+1}
  {+}2 h \hat \sigma^z_n\big) ~,
\end{equation}
where $(\hat\sigma^x_n, \hat\sigma^y_n, \hat\sigma^z_n)$ are the
Pauli matrices for the spin at site $n$, $J$ is the homogeneous
coupling, and $h$ is the magnetic field. The $2M$ lattice sites are
labelled by the half-integer index
$n=-M+\frac{1}{2},...M-\frac{1}{2}$. Correspondingly, the lattice
bonds are labelled by the integer index $b=-M+1,...,M$, with
$b=n\,{+}\,1/2$ indicating the bond between sites $n-1$ and $n$. This
notation allows the reflection symmetry with respect to the impurity
bond to emerge more clearly in many of the following equations
involving the correlation functions which, on the other hand, refer
to lattice sites. The enforcement of the PBC conditions
${\hat{\vec{\sigma}}}_{M+n}\,{=}\,{\hat{\vec{\sigma}}}_{-M+n}$ makes
Eq.~\eqref{e.hamilton0} the Hamiltonian of a ring.

\begin{figure}[b!]
\includegraphics[width=60mm]{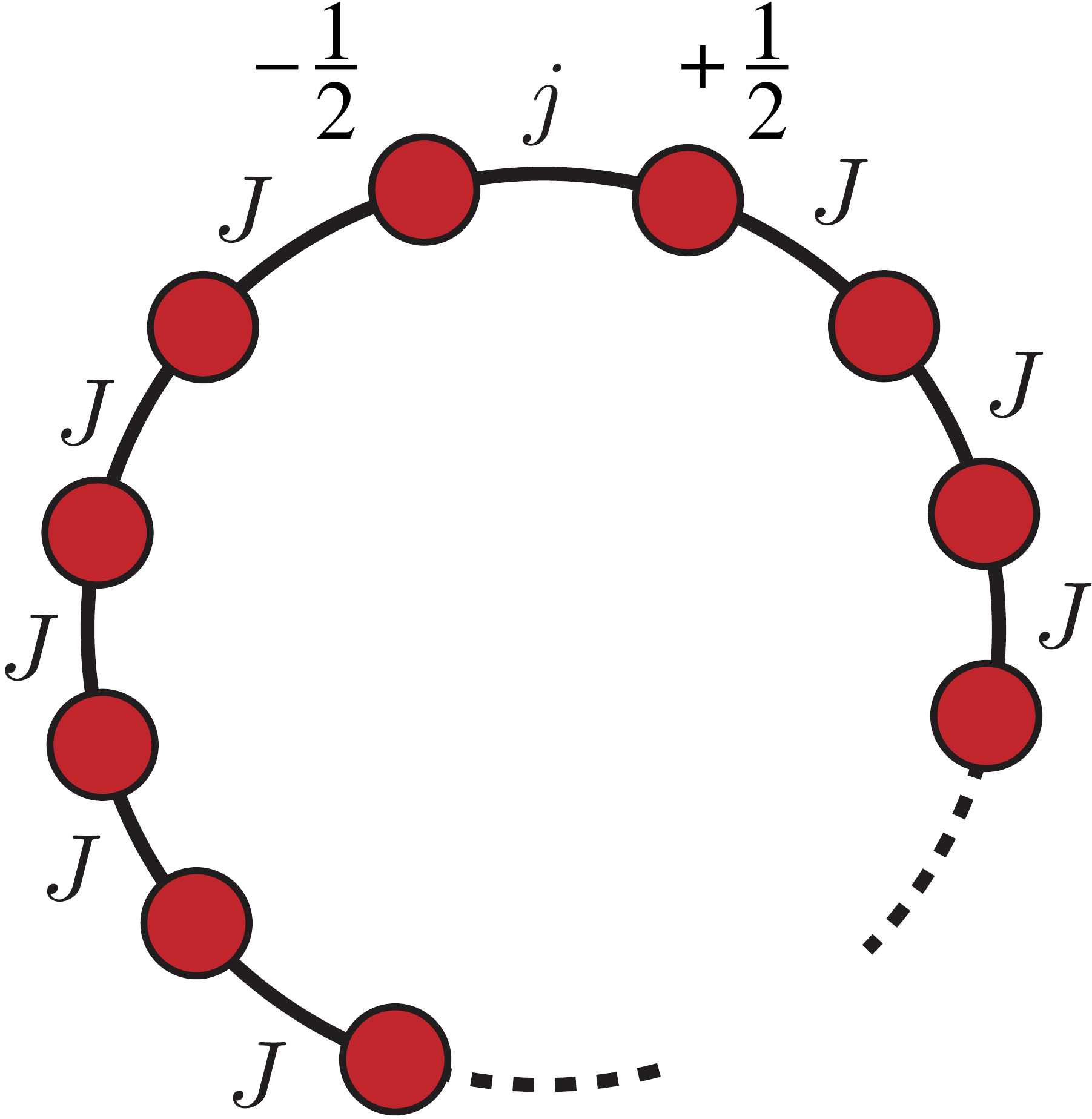}
\caption{(Color online) A ring of interacting spin-$1/2$ particles,
all coupled through an XX model, includes a bond defect: while all
spin pairs $(n,n+1)$ with $n{\neq}-1/2$ are mutually interacting with
strength $J$, the pair $(-1/2,1/2)$ experiences the strength $j$. The
spins are all subjected to a homogeneous magnetic field $h$.}
\label{f.modello}
\end{figure}

We introduce a single bond impurity (BI)
by varying the exchange integral that generates the bond $b{=}0$, i.e. the interaction strength between the two spins on sites
${n{=}{-}\frac12}$ and $n{=}\frac12$ (which we will refer to as the impurity spins). This implies adding the term
\begin{equation}
 \hat{\cal H}_{\rm{I}} {=} \frac{J{-}j}2\big(
 \hat\sigma^x_{-\frac12} \hat\sigma^x_{\frac12}
 + \hat\sigma^y_{-\frac12} \hat\sigma^y_{\frac12}\big)
\label{e.hamilton-imp}
\end{equation}
to the translation-invariant Hamiltonian in
Eq.~\eqref{e.hamilton0}. From now on, we assume $J{=}1$ as the
energy unit. The resulting system
$\hat{\cal{H}}{=}\hat{\cal{H}}_0\,{+}\,\hat{\cal{H}}_{\rm{I}}$ is
illustrated in Fig.~\ref{f.modello}, where $j$ gives the coupling
strength and $j{=}1$ ($j{=}0$) corresponds to the well-known
$2M$-PBC ($2M$-OBC) spin chains~\cite{LiebSM1961}. For every value
different from the two cases above, we diagonalize the Hamiltonian
as follows.

The total Hamiltonian $\hat{\cal H}_0+\hat{\cal H}_{\rm{I}}$ can be
mapped via the Jordan-Wigner transformation~\cite{LiebSM1961} into
\begin{equation}
\label{e.hamiltonf}
 \hat{\cal H}{=}{-}\mkern-18mu \sum_{n{=}{-}M{+}\frac12}^{M{-}\frac12}
\mkern-18mu  (\hat c^\dagger_{n+1}\hat c_n{+}h.c.
 {+} 2 h \hat c^\dagger_n \hat c_n)
 {-}(j{-}1)(\hat c^\dagger_{\frac12} \hat c_{-\frac12}{+}h.c.),
\end{equation}
where $\{c_n,c^{\dagger}_n\}$ are the fermionic destruction and
creation operators. As translation symmetry is broken for $j{\ne}1$,
a Fourier transform does not diagonalize Eq.~\eqref{e.hamiltonf}. It
is nevertheless possible to solve the model analytically by making
use of a Green function approach~\cite{Economoubook,Pury1991}. The
key steps of this procedure are outlined in Appendix~\ref{a.diag} and
the diagonalized Hamiltonian in the thermodynamic limit finally reads
\begin{equation}
\label{e.hamiltonfin}
 \hat{\cal H}\,{=}\,\int_{-\pi}^{\pi}\frac{dk}{2\pi}~
 E_k~\hat\zeta^{\dagger}_k\hat\zeta_k
 +E_+~\hat\zeta^{\dagger}_+ \hat\zeta_+
 +E_-~\hat\zeta^{\dagger}_- \hat\zeta_- ~.
\end{equation}
The first term represents the intra-band contributions  and we have
introduced the operators
\begin{equation}\label{E.modes}
 \hat\zeta_k=\frac1{\sqrt{2M}}
                   \sum_ne^{-ikn}(1{+}f_{kn})\,\hat{c}_n~,
\end{equation}
which annihilate fermions with energy $E_k=2(\cos{k}-h)$.
Here, the functions $f_{kn}$ account for the spatial distortion of
the intra-band excitations as
\begin{equation}\label{E.distorsion}
 f_{kn} {=} \left\{\begin{aligned}
 &\frac{i(j^2{-}1)e^{2ikn}}
                     {2\sin|k|{-}i(j^2{-}1)e^{i|k|}}
 ~~~~~~~~~~{\rm if}~~ kn>0,~~
\\
 &\frac{2(j{-}1)\sin|k|{+}i(j^2{-}1)e^{i|k|}}
                     {2\sin|k|{-}i(j^2{-}1)e^{i|k|}}
  ~~~{\rm if}~~ kn<0,~~
\end{aligned}\right.
\end{equation}
Such distortion is evidently due to the BI ($f_{kn}{=}0$ for
$j{=}1$), and is responsible for the oscillations observed in the
correlations, as discussed in the following Section. The second term
of Eq.~\eqref{e.hamiltonfin} accounts for two discrete-energy
eigenstates $E_\pm$ which appear only for $j\,{>}\,1$: their energies
are $E_\pm\,{=}\,-2h\pm(j{+}1/j)$, above and below the band,
respectively. They correspond to excitations that, once expressed in
terms of direct lattice-site fermionic operators, take the form
\begin{equation}\label{E.locstates}
 \hat \zeta_{\pm}=\sqrt{\sinh{q}}
   \sum_n(\pm)^{n+\frac12}e^{-q|n|}\,\hat c_n
\end{equation}
with $q=\ln{j}$ being the reciprocal of the localization length.

Let us now compare the behavior of the system in the two extreme
cases of $j\,{=}\,0$ and $j\,{\to}\,\infty$.
First, one can easily see that for $kn\,{<}\,0$ in both cases one has
$f_{kn}\,{=}\,-1$, namely the impurity acts as a purely reflective
barrier yielding complete backscattering.
On the other hand, for $kn\,{>}\,0$  the distortions of the in-band
excitations in the two limits read
$f_{kn}\,{=}\,-e^{i(2kn{\pm}|k|)}$, respectively. It follows that for
$j\,{\to}\,\infty$ the distortion at the impurity sites is
$f_{k,\frac12}=f_{k,-\frac12}=-1$, meaning that these sites
completely decouple from the rest of the system, their state being
exclusively determined by the two, now completely localized,
out-of-band states
$\ket{E_{\pm}}\,{=}\,\frac1{\sqrt2}\big({c^{\dagger}_{\frac12}\mp
c^{\dagger}_{-\frac12}}\big)\ket{0}$: as they do not take part in the
dynamics, the spins at sites $n=\pm 3/2$ take the role of head and
tail of a segment of length $2M{-}2$. Of course, for $j\,{=}\,0$ the
resulting segment has length $2M$. This argument suggests that one BI
can indeed change the boundary conditions from PBC to OBC.  In other
terms, a segment can be obtained not only by actually cutting the
ring ($j=0$), but also by making the interaction between the spins
sitting at sites $n\,{=}\,\pm1/2$ strong enough with respect to the
coupling between all the other nearest-neighbor spins
($j\,{\gg}\,1$), as to effectively decouple them from the rest of the
system.

In the next Section we further explore this idea in the case
$M\to\infty$, where the availability of the analytical results
presented here allows us, through a straightforward application of
Wick's theorem~\cite{LiebSM1961}, to exactly evaluate two-points
correlations functions, concurrence and quantum
discord~\cite{discord1,discord2,discordreview}. We focus on the
possibility that the efficiency of the ring-cutting mechanism
described above holds for moderately large values of $j$.

\section{Effective ring-cutting mechanism: study of the two-point functions}
\label{S.cutting-2points}

\begin{figure}
\includegraphics[width=\columnwidth]{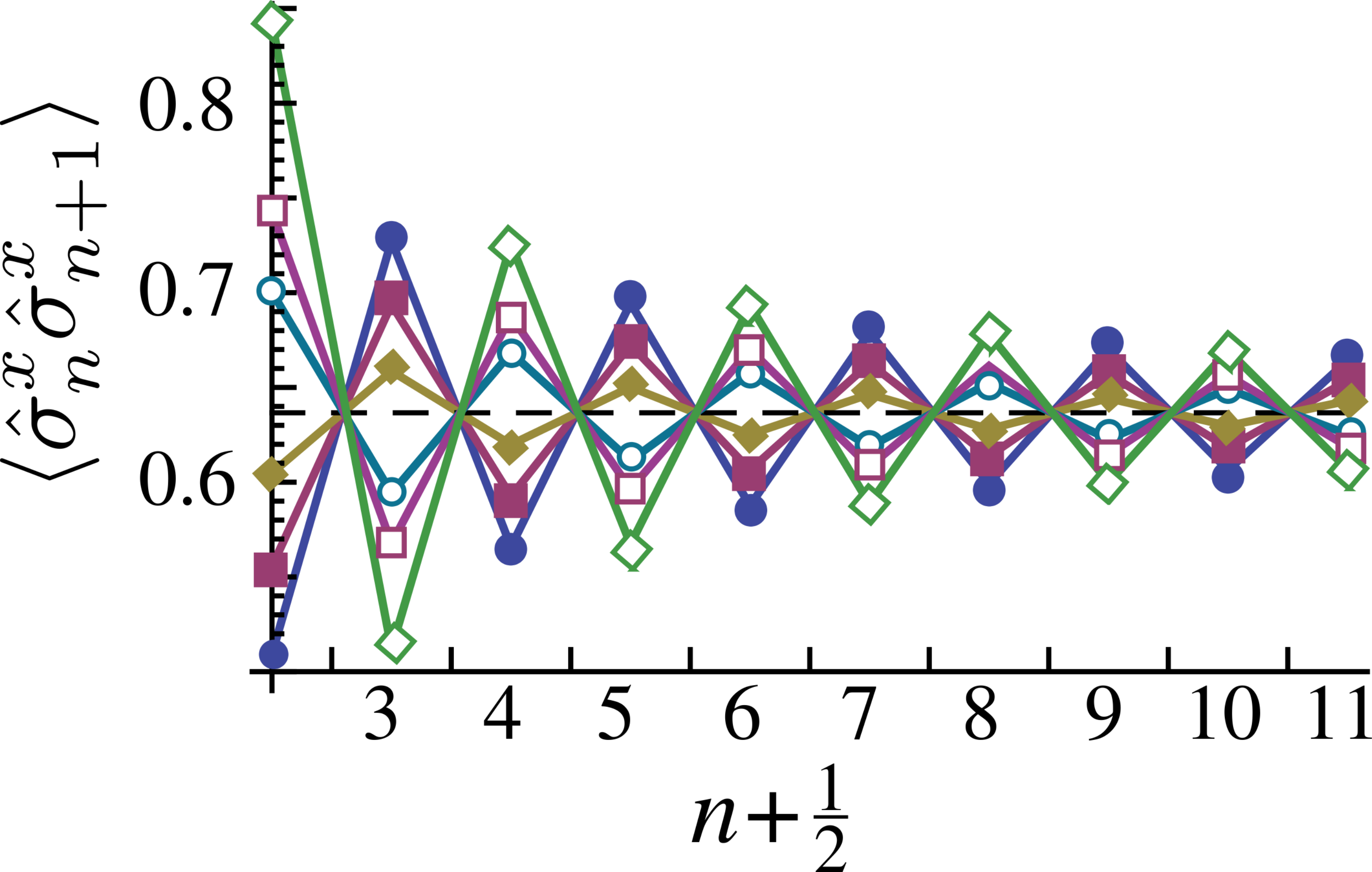}
\\
\includegraphics[width=\columnwidth]{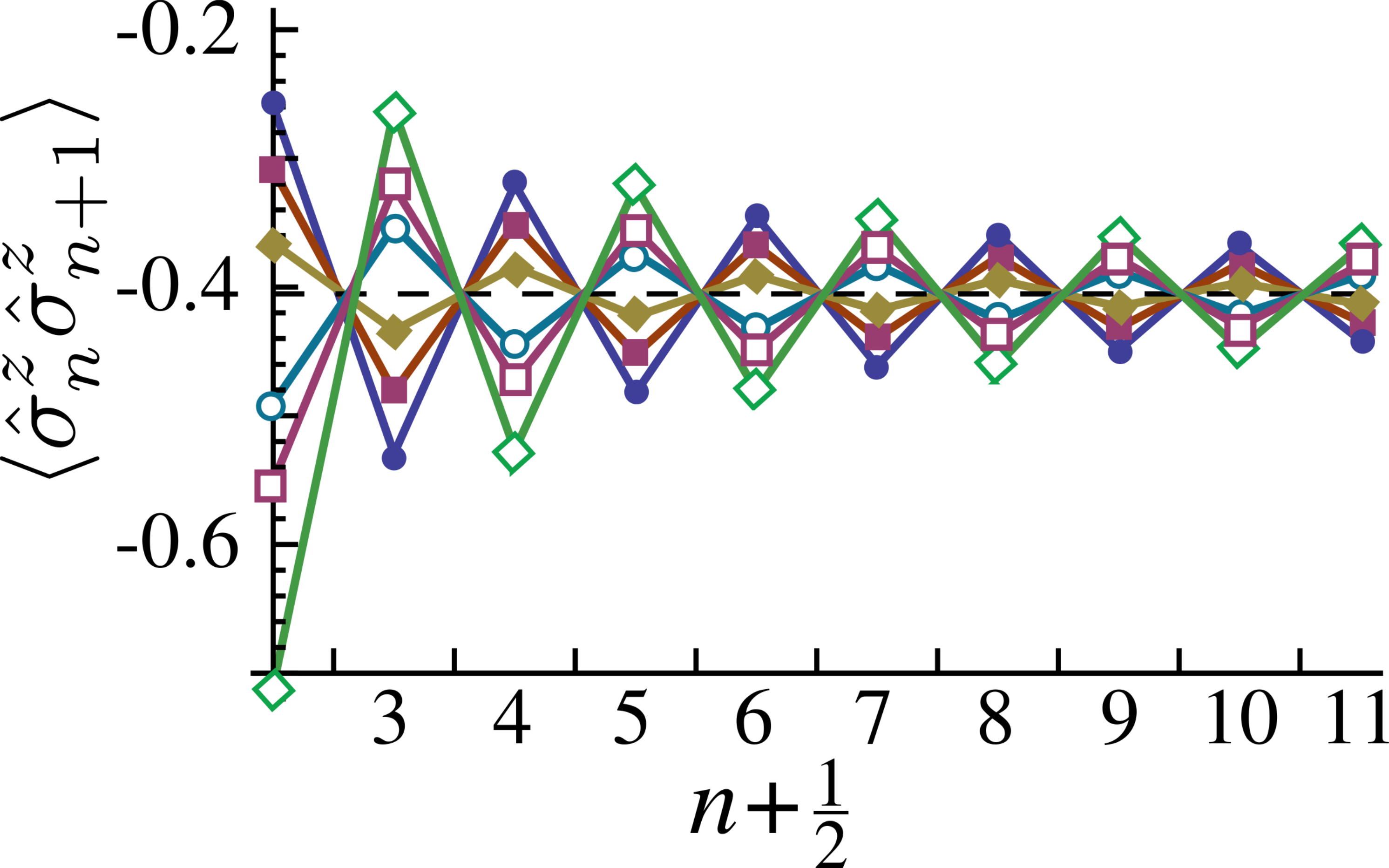}
\caption{(Color online)
Correlators $\valmed{\hat\sigma_n^x\hat\sigma_{n+1}^x}$ (top) and
$\valmed{\hat\sigma_n^z\hat\sigma_{n+1}^z}$ (bottom) for $j{=}0, 0.5,
0.8, 1.5, 2, 11$ (corresponding to increasing absolute values for
$n{+}1/2$ even). The straight lines correspond to the correlators in
the PBC case, $j\,{=}\,1$. For $j{=}11$ the data are indistinguishable
from the OBC limit.}
\label{F.xxzzoscillh0}
\end{figure}

\begin{figure}[b!]
\includegraphics[height=80mm,angle=90]{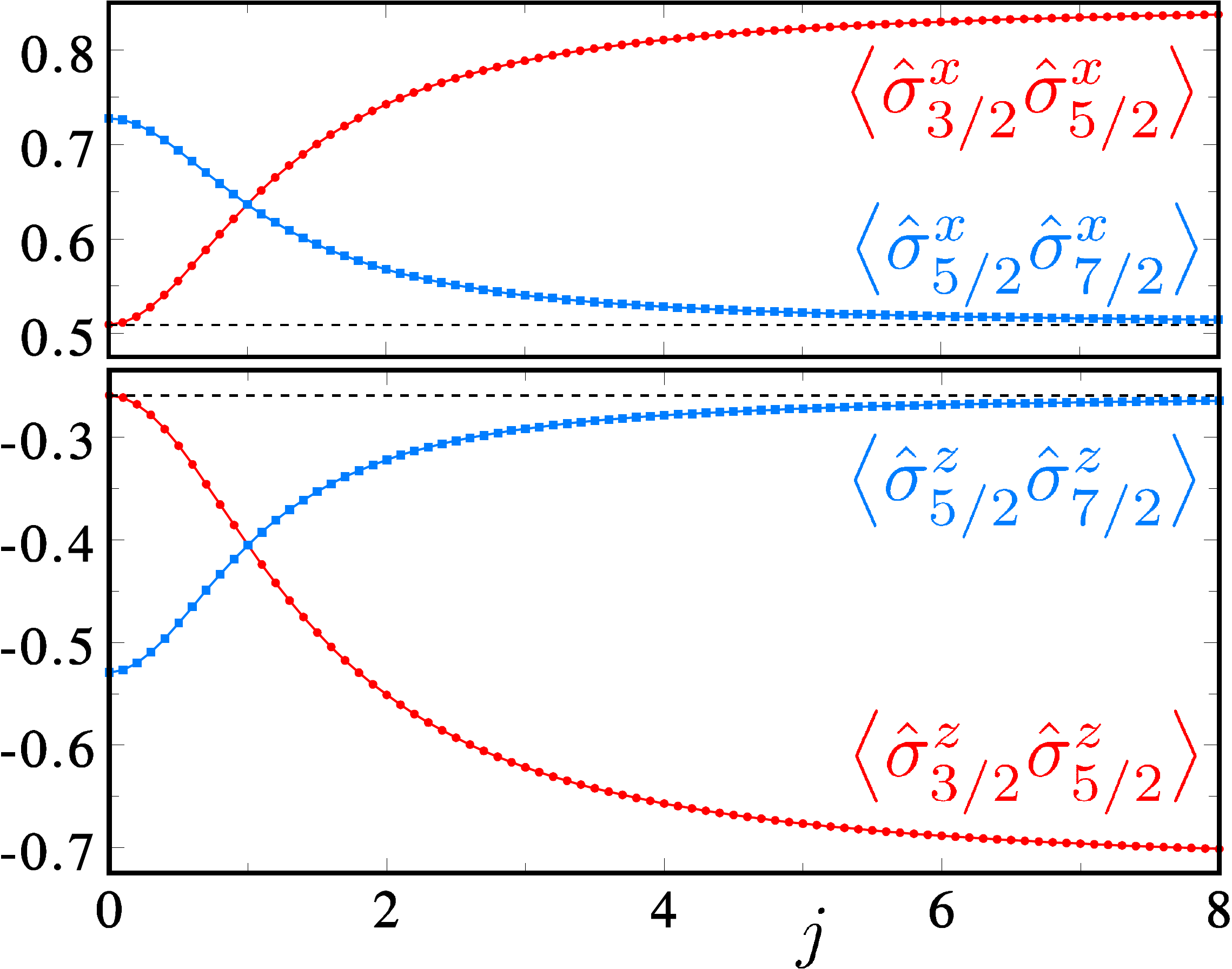}
\caption{(Color online) The nearest-neighbor correlation functions
$\valmed{\hat\sigma_n^x\hat\sigma_{n+1}^x}$ and
$\valmed{\hat\sigma_n^z\hat\sigma_{n+1}^z}$ corresponding to the
second and third bond after the defect, vs $j$. The $xx$ ($zz$)
correlators take positive (negative) values; their absolute value
increases (decreases) with $j$ for $n\,{=}\,3/2$ ($n\,{=}\,5/2$). The
dashed lines show that the third-bond correlators at $j\to\infty$
behave as the second-bond correlators of the open chain, i.e. $j{=}0$.}
\label{F.corr12e23}
\end{figure}

In this Section we study the effects of the BI on two-point
functions, i.e. quantities relative to spin pairs. As far as we only
consider pairs of nearest-neighbor spins, such quantities can be
labelled by the integer bond-index $b$ representing the distance in
lattice spacings from the BI, according to $O_{n,n+1}=O_b$ with
$b=n{+}{1}/{2}$. We first analyze the nearest-neighbor magnetic
correlations
$g^{\alpha,\alpha}_b{\equiv}\valmed{\sigma^\alpha_n\sigma^\alpha_{n+1}}$
($\alpha=x,z$). For any $j{\neq} 1$, Friedel-like oscillations appear
and induce a spatial modulation of the correlations with periodicity
$p{=}{\pi}/{k_F}$, where $k_F{=}\cos^{-1} h$ is the Fermi momentum.
In Figs.~\ref{F.xxzzoscillh0} we consider $h=0$, corresponding to
$p=2$, and study $g^{\alpha,\alpha}_b$ against the value of $b$ for
various choices of $j$. The presence of the BI modifies the strength
of correlations and the following relations ($b$ is an integer)
clearly emerge
\begin{equation}
\begin{aligned}
&|g^{\alpha,\alpha}_{2b}(j{<}1)|<|g^{\alpha,\alpha}(j{=}1)|~{<}~|g^{\alpha,\alpha}_{2b}(j{>}1)|,\\
&|g^{\alpha,\alpha}_{2b{+}1}(j{<}1)|>|g^{\alpha,\alpha}(j{=}1)|~{>}~|g^{\alpha,\alpha}_{2b{+}1}(j{>}1)|,
\end{aligned}
\end{equation}
where the bond-index dependence is omitted for $j{=}1$, as in the
uniform case PBC guarantee translation invariance. From the above
inequalities we deduce that the results corresponding to the  limit
$j{\to}\infty$ cannot be possibly related with the behavior of the
segment obtained by an actual cut, i.e. what is found by setting
$j{=}0$. Indeed, $g^{\alpha,\alpha}_b(\infty)$ is in general
different from $g^{\alpha,\alpha}_b(0)$. In fact, as already
mentioned at the end of the above Section, we expect the
$j{\to}\infty$ limit to reproduce the behavior of a segment with head
and tail at $n{=}\pm{3}/{2}$, i.e. $b{=}\pm2$. Therefore, in all
those cases for which the actual value of $M$ is not relevant, such
as in the thermodynamic limit considered here, the meaningful
comparison to be performed involves $g^{\alpha,\alpha}_b(j{=}0)$ and
$g^{\alpha,\alpha}_{b{+}1}(j{\to}\infty)$. In order to quantitatively
check to what extent a model with large $j$ can be actually
considered to behave as a segment, in Fig.~\ref{F.corr12e23} we
compare $g^{\alpha,\alpha}_2$ and $g^{\alpha,\alpha}_3$ for
increasing values of $j$. Clearly, the correlations along $x$ and $z$
almost match the values corresponding to a true segment already for
$j>8$, confirming that an effective ring-cutting mechanism takes
place.

In order to provide an all-round characterization of our proposal,
we now complement the analysis performed above by addressing the
leakage of information out of head and tail of the segment
effectively obtained by increasing $j$. We quantify the extent of
such leakage by addressing the values taken by both classical
correlations (CC) and quantum discord
(QD)~\cite{discord1,discord2,discordreview,nota} {\it across} the
impurity, i.e., between two spins sitting on opposite sides with
respect to the BI, normalized by their respective values for $j{=}1$.

\begin{figure}
\includegraphics[height=80mm,angle=90]{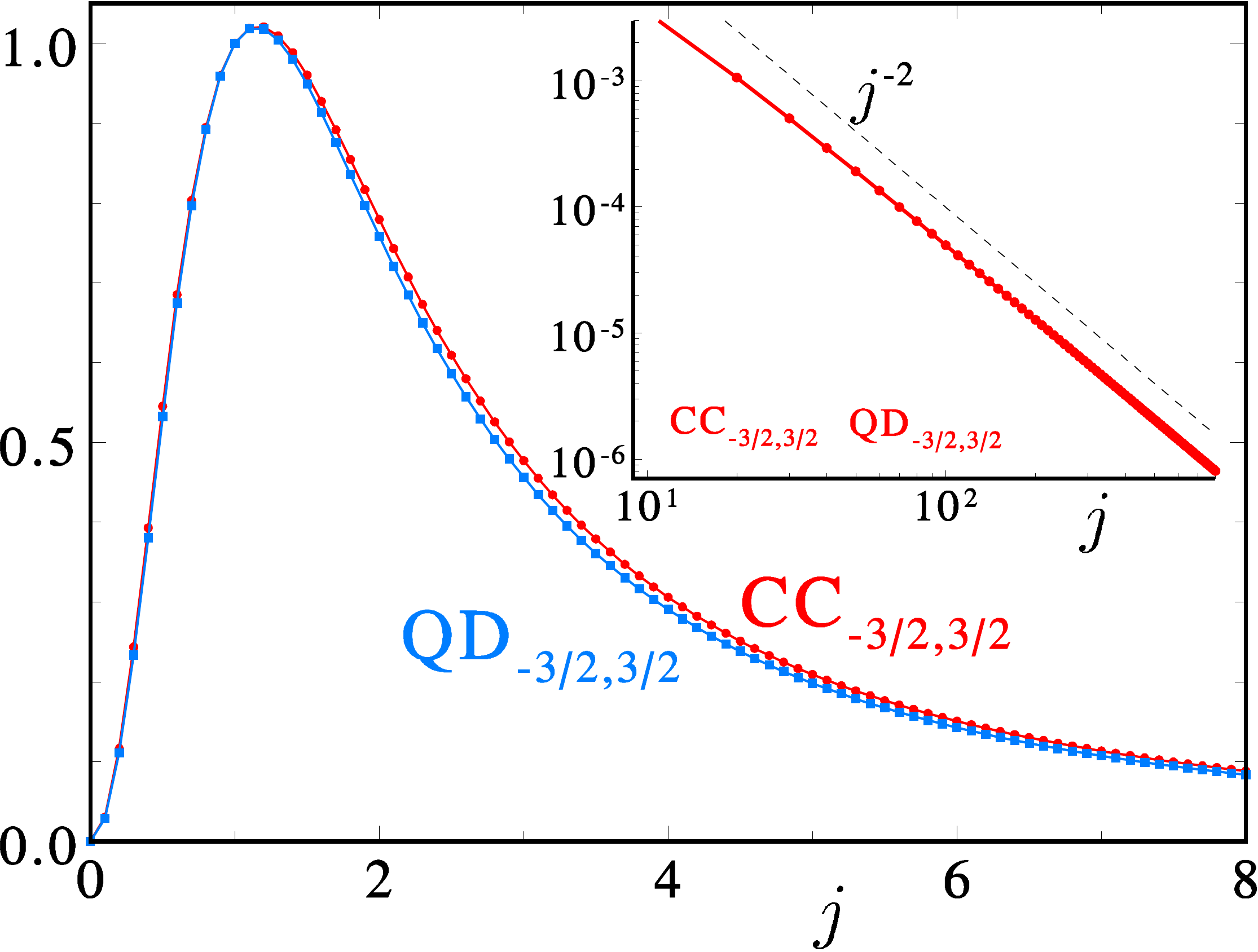}
\caption{(Color online) QD and CC (normalized with respect
to the $j\,{=}\,1$ value) plotted vs $j$ for the two spins at sites
$\pm3/2$, i.e., sitting at opposite sides of the impurity. Cutting
the chain affects both quantum and classical correlations in an
essentially identical way. Inset: log-log plot of QD and CC (here
indistinguishable) vs $j$, showing that they obey a $j^{-2}$ scaling
law, which is in fact independent of the distance of the sites.}
\label{F.decaywithinset}
\end{figure}

The results corresponding to considering the spins at sites
$n={\pm}{3}/{2}$ are shown in Fig.~\ref{F.decaywithinset}. Both CC
and QD across the BI are non-monotonic functions of the strength
$j$. For small values of $j$, both rapidly grow. On the other
hand,  the range $j\gg1$ corresponds to the monotonic decrease of
all forms of correlations, thus demonstrating that the ring is
effectively cut. Remarkably, for $j\,{\gtrsim}\,1$ ,CC and QD are
larger than their value at $j\,{=}\,1$. This is due to the spread
of the localized state over these sites, yielding an enhancement
similar to that reported in Refs.~\onlinecite{OsendaHK2003,
ApollaroCFPV2008,PlastinaA2007}, to which we refer for a detailed
discussion. CC and QD behave in very similar ways, decaying
asymptotically, for $j\gg1$, as $j^{-2}$ (cf. the inset of
Fig.~\ref{F.decaywithinset}). This power-law decay stems from the
behavior of the magnetic correlations. In fact, these enter both
the expression of the concurrence (cf. Eq.~\eqref{E.Conc} below)
and those of QD and CC (which are not reported here as too lengthy
to be informative). In particular, by considering
Eqs.~\eqref{E.modes} and~\eqref{E.distorsion} in the $j\gg 1$
limit, and evaluating by standard methods the magnetic correlation
functions (as done, for instance, in Ref.~\cite{LiebSM1961}), we
find that $\valmed{\hat\sigma_n^x\hat\sigma_{m}^x}={\cal
O}\left(j^{-\left(\left|n\right|+\left|m\right|\right)}\right)$,
whereas $\valmed{\hat\sigma_n^z\hat\sigma_{m}^z}={\cal
O}\left(j^{-2}\right)$, regardless of the relative distance
between the spins. As a consequence, the scaling law $j^{-2}$
reported in the inset of Fig.~\ref{F.decaywithinset} originates
from the correlations along the $z$-axis and is thus independent
of the site-separation. On the contrary, the correlation functions
along the $x$-axis shown in Fig.~\ref{F.corr12e23} (a) do depend
on the distance, as reported above.

\begin{figure}[h]
 \centering

   {\includegraphics[width=0.8\columnwidth]{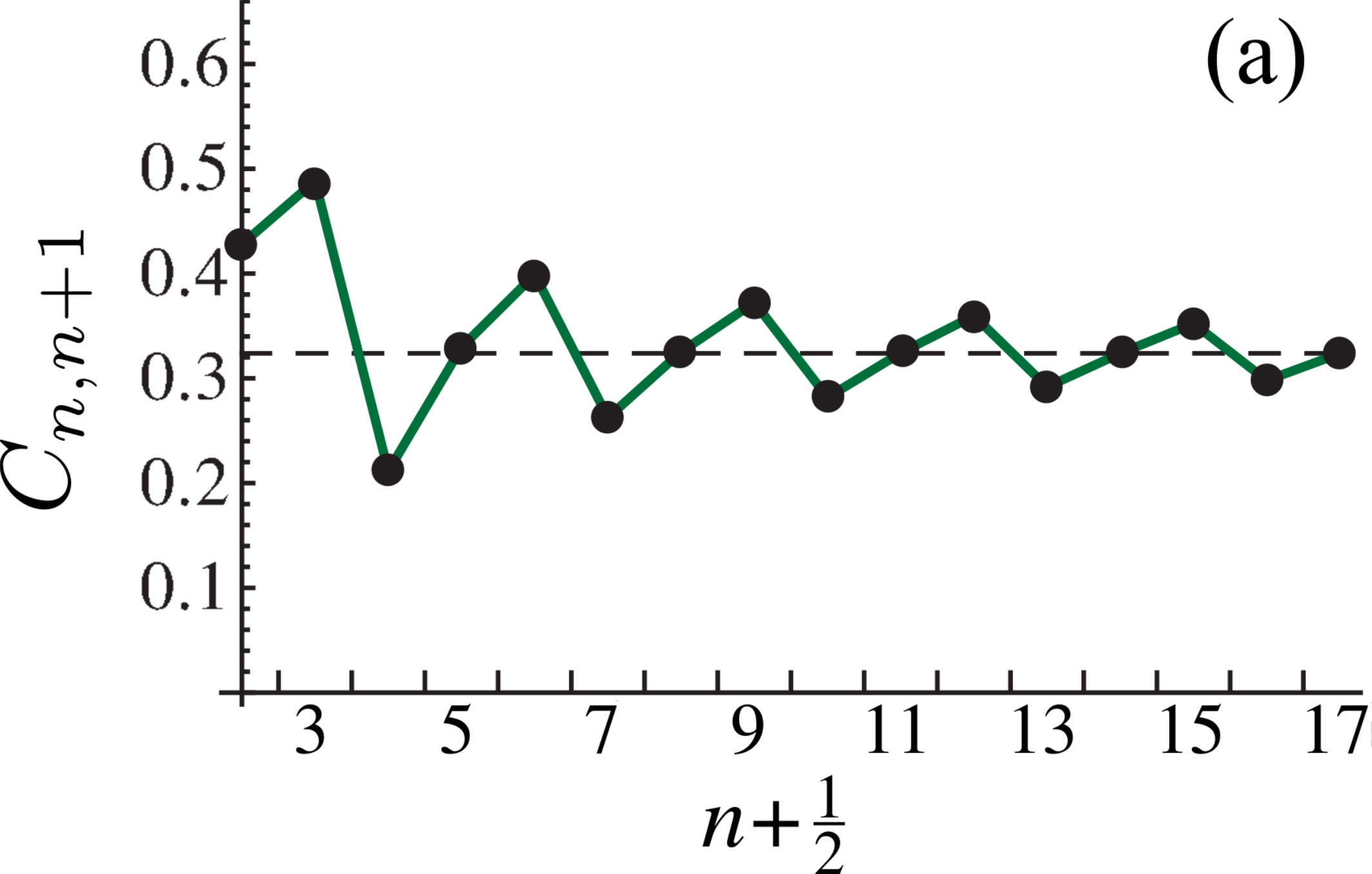}}\\
{\includegraphics[width=0.8\columnwidth]{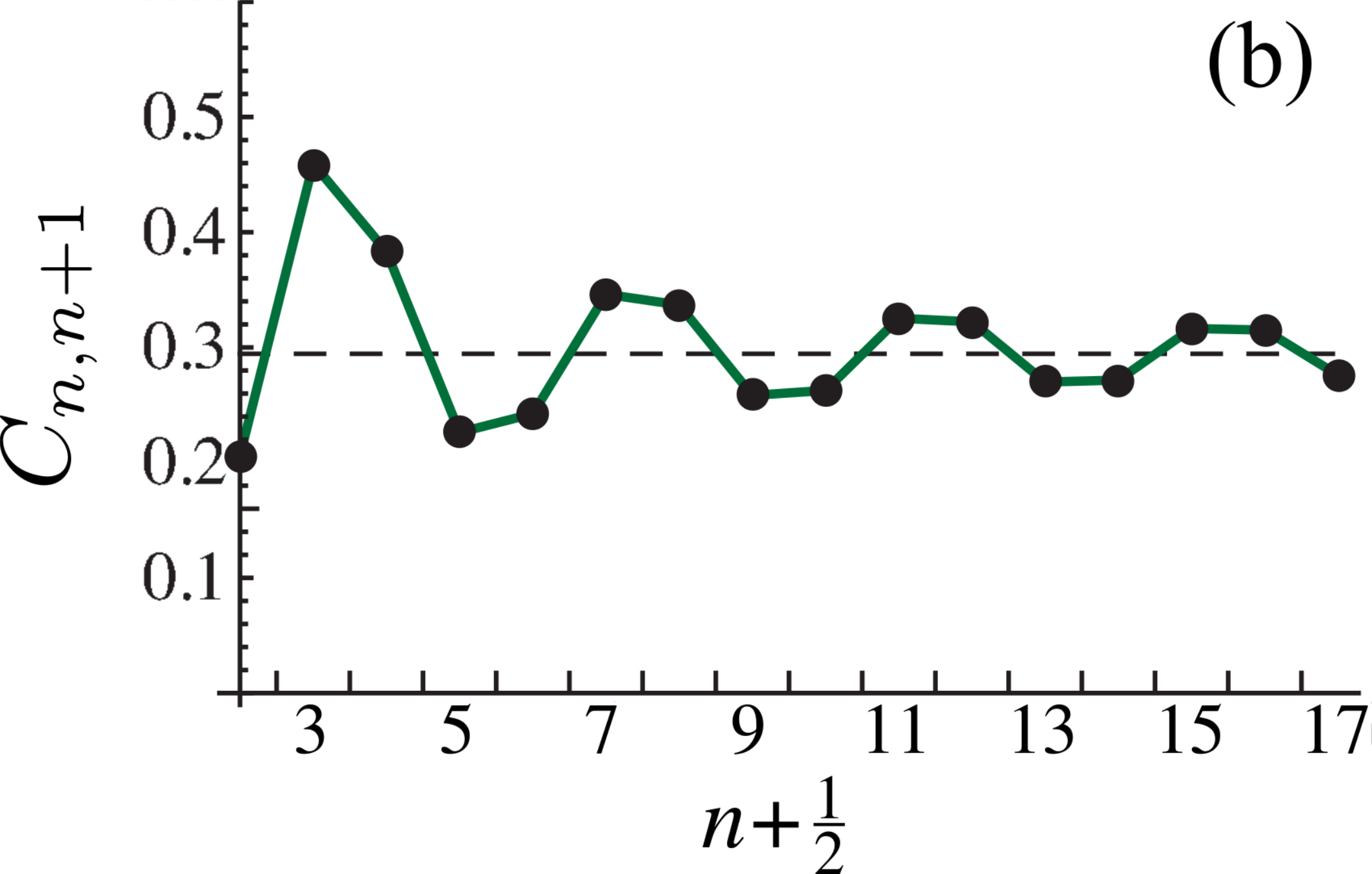}}
\caption{Nearest-neighbor concurrence $C_{n,n+1}$ for $j\,{=}\,6$ vs
the bond index $n{+}\frac12$. Panels (a) and (b)  are for
$h\,{=}\,0.5,\frac{1}{\sqrt{2}}$ respectively. The straight dashed
line shows the value of the concurrence at $j\,{=}\,1$. The magnetic
field sets the periodicity of the one- and two-points spin
correlators, which enter the concurrence, to $p\,{=}\, 3, 4$
respectively.}
\label{F.chvar}
\end{figure}

We conclude this Section by briefly discussing how, by tuning the
intensity of the impurity strength, it is possible to exploit the
Friedel oscillations in order to spatially modulate the
concurrence~\cite{wootters} between neighboring spins. In
Fig.~\ref{F.chvar} we show the nearest-neighbor concurrence for
$j{=}6$ at different values of $h$. Analytically, the concurrence
$C_{n,m}$ depends on the magnetic correlation functions
as~\cite{Fubinietal2006}
\begin{equation}\label{E.Conc}
C_{n,m}\,{=}\,\max{[0,\valmed{\hat\sigma^x_n{\otimes}
\hat\sigma^x_m}{-}\frac12\sqrt{(S^{zz}_{nm})^2{-}(s^{zz}_{nm})^2}]}
\end{equation}
with
$S^{zz}_{nm}{=}1{\pm}\valmed{\hat\sigma^z_n{\otimes}\hat\sigma^z_m}$
and
$s^{zz}_{nm}{=}\valmed{\hat\sigma^z_n}{+}\valmed{\hat\sigma^z_m}$.
The values of $C_{n,m}$ achieved in our system are the same as those
of an open-boundary spin chain in the presence of a strong magnetic
field on a single spin~\cite{SonAPV09,ApollaroCFPV2008}. Moreover, we
notice the presence of a periodic spatial modulation (with respect to
the value of concurrence achieved for PBC), determined by the
periodicity $p={\pi}/{\cos^{-1} h}$ of the Friedel oscillations, as
reported also for different impurity types in
Refs.~\onlinecite{OsendaHK2003,ApollaroCFPV2008.2}.

\section{Effective ring-cutting mechanism: analysis of the state fidelity}
\label{S.fidelity}

In order to further verify the efficiency of the proposed mechanism,
we now take a different point of view and consider a global figure of
merit  from which we can obtain indications on the similarity between
the state of the cut ring and that of a true segment. As a
description of the state of the former, we choose the reduced density
matrix
$\rho{=}\Tr_{n{=}{\pm\frac{1}{2}}}\left[\ket{\Omega}\!\!\bra{\Omega}\right]$
of a $2(M-1)$ spin system where the impurity spins have been traced out of the ring. As for the state of a segment, which embodies our target state, we take the pure state $\ket{\Sigma}$ of a system of $2(M-1)$ spins with OBC.
As a measure of closeness between two quantum states we use the
quantum fidelity~\cite{Josza94} $\mathcal
F\!\left(\ket{\Sigma}\!,\!\rho\right) {=}
\valmed{\Sigma\vert\rho\vert\Sigma}$.

The ground state of a free-fermion model such as the one in
Eq.~\eqref{e.hamiltonf} is given by
\begin{align}
    \ket{\Omega} {=}\!\! \prod_{k{:}E_k{<}0}\!\!\! \hat\zeta_k^\dagger \ket 0,
    \label{e.diracsea}
\end{align}
for which all the negative-energy eigenstates  up to the Fermi energy $E_{k_F}{=}0$ are occupied by a fermionic quasi-particle, whereas positive-energy levels are empty.
As a consequence, states with a different number of fermions yield zero fidelity.
As the number of fermions in the Dirac sea is given by the intensity of the magnetic field $h$, which sets the Fermi momentum,  we will compare the actual state of the cut ring with a target state for the same value of the applied magnetic field.
A somewhat lengthy but otherwise straightforward calculation based on the use of Wick's theorem shows that $\mathcal F$ depends on the submatrices
of the transformation mapping the
real-space fermions $\hat c_n$ to those diagonalizing the Hamiltonian in the case of Eq.~\eqref{e.hamiltonfin} (the target model)
for $n{=}{-}M{+}1/2,\dots,M{-}1/2$ and $k<k_F$. Some details of this derivation
are sketched in Appendix~\ref{a.fid}.
\begin{figure}
 \centering
   \includegraphics[width=0.9\columnwidth]{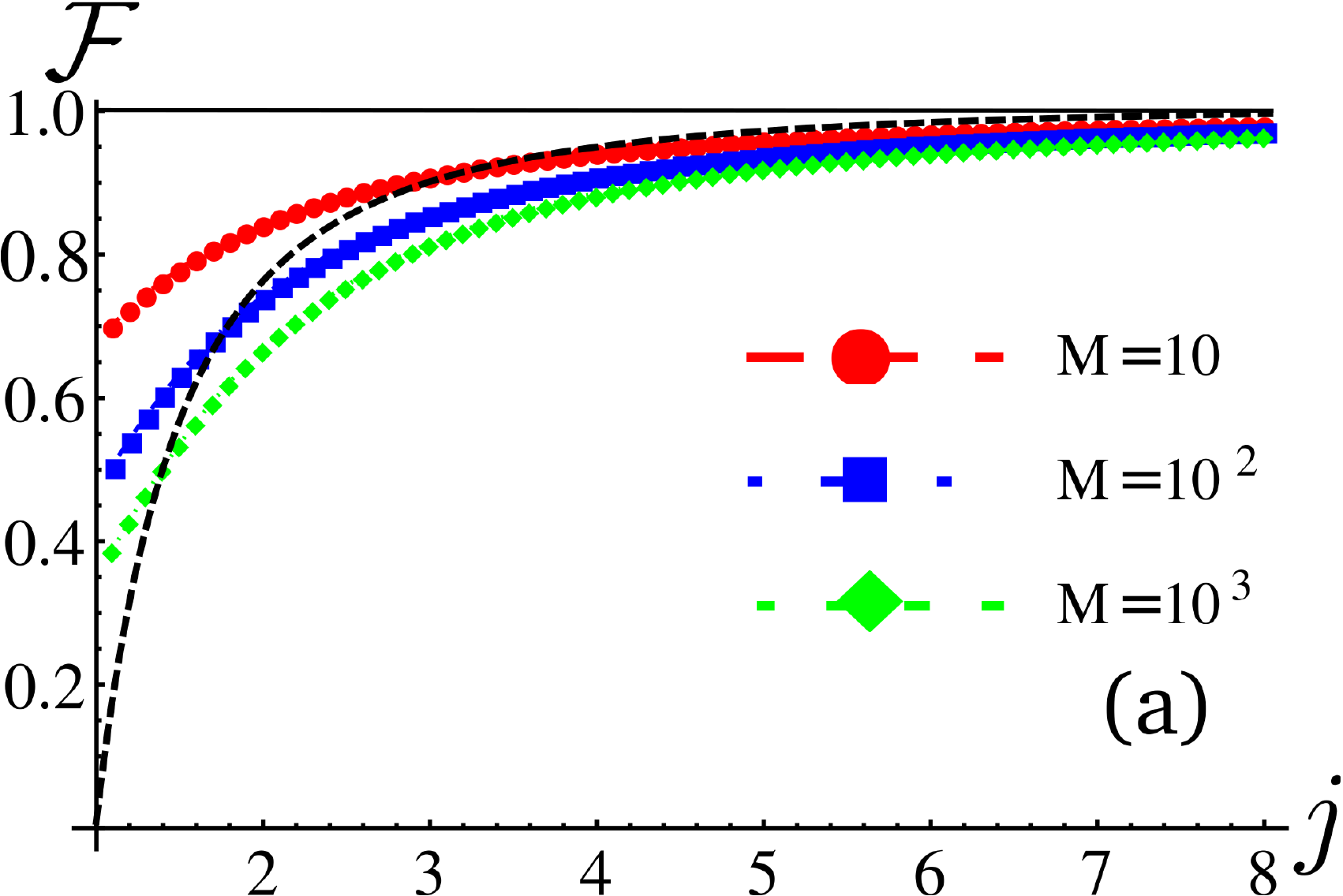}\\
   \includegraphics[width=0.9\columnwidth]{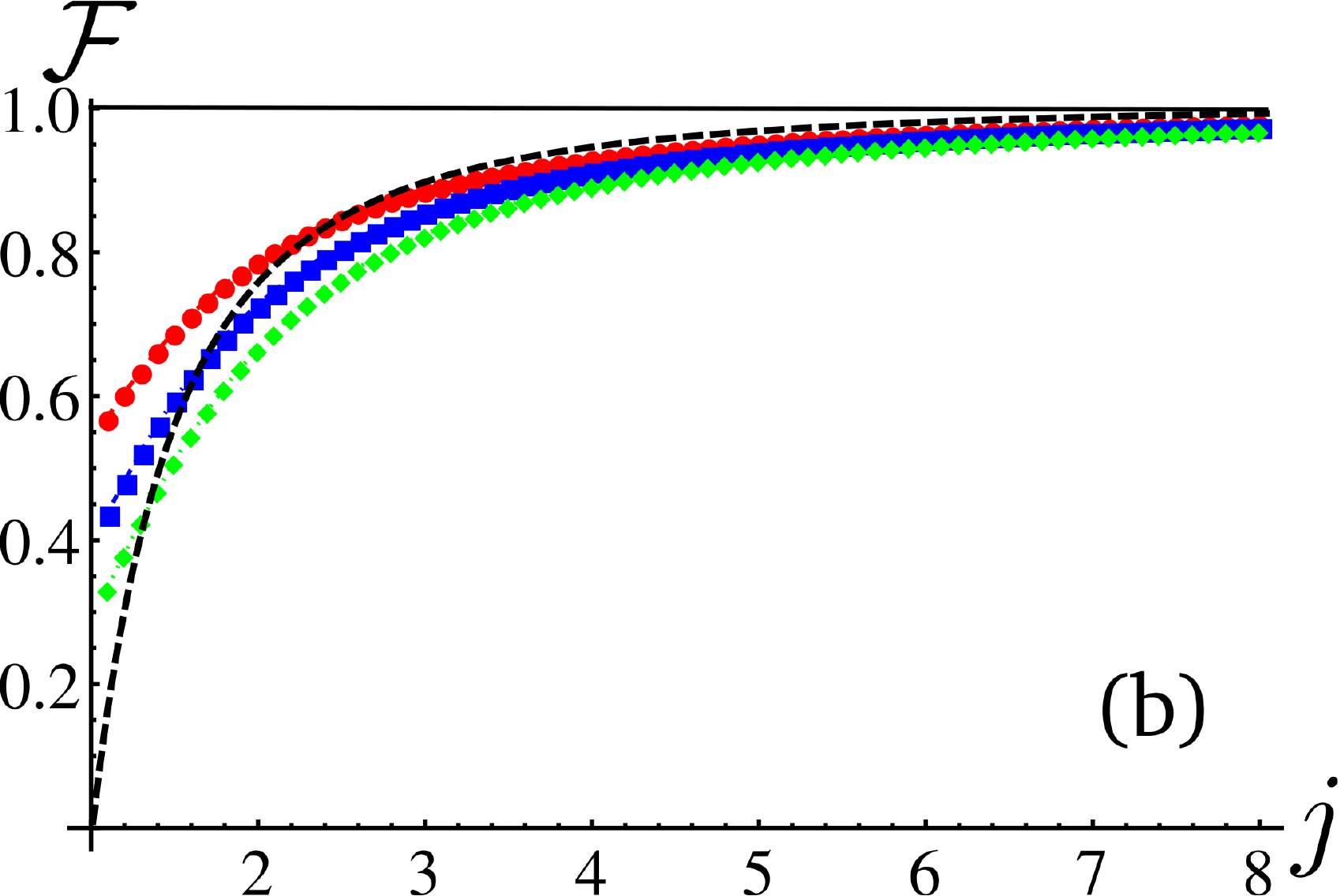}\\
   \includegraphics[width=0.9\columnwidth]{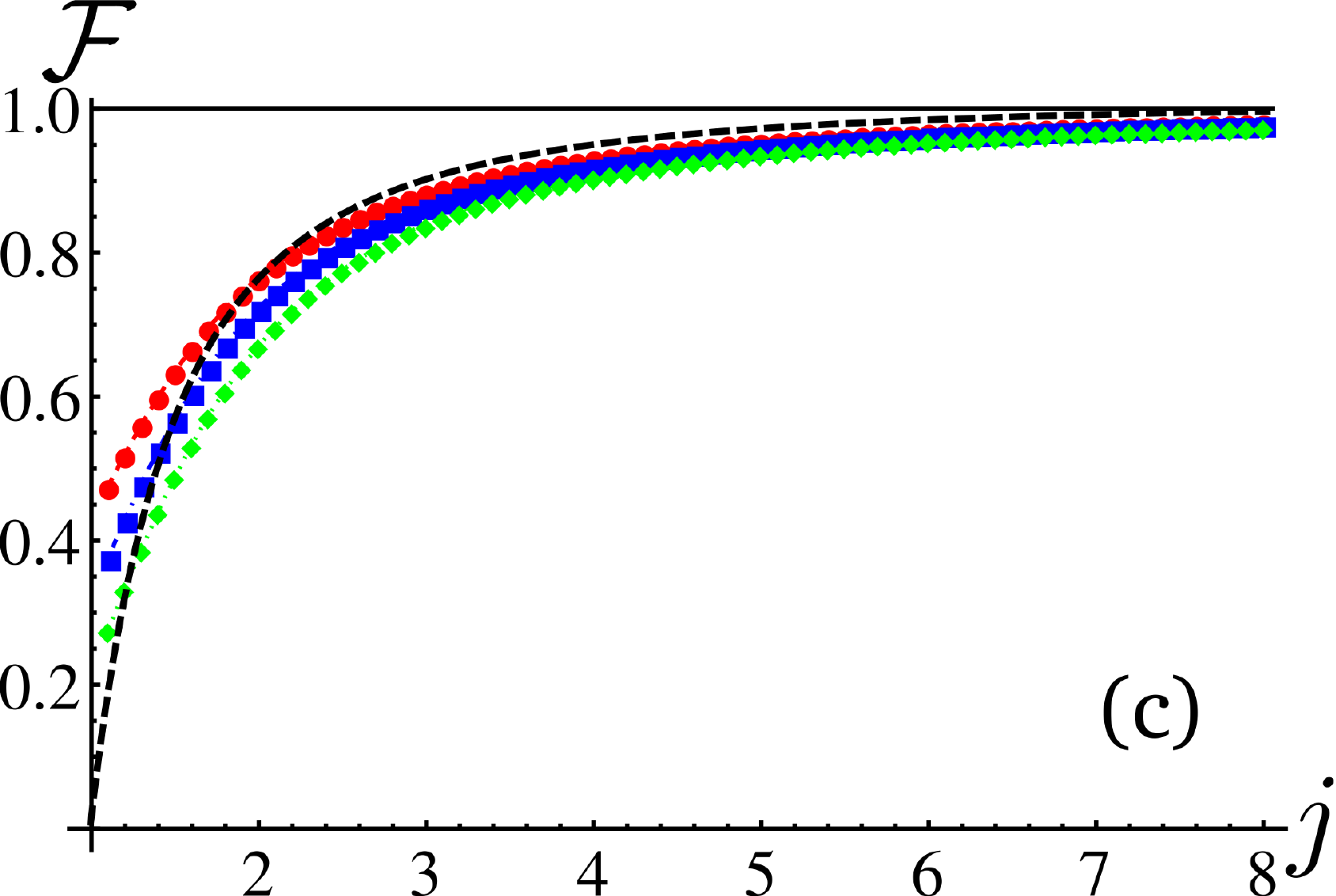}\\
   \includegraphics[width=0.9\columnwidth]{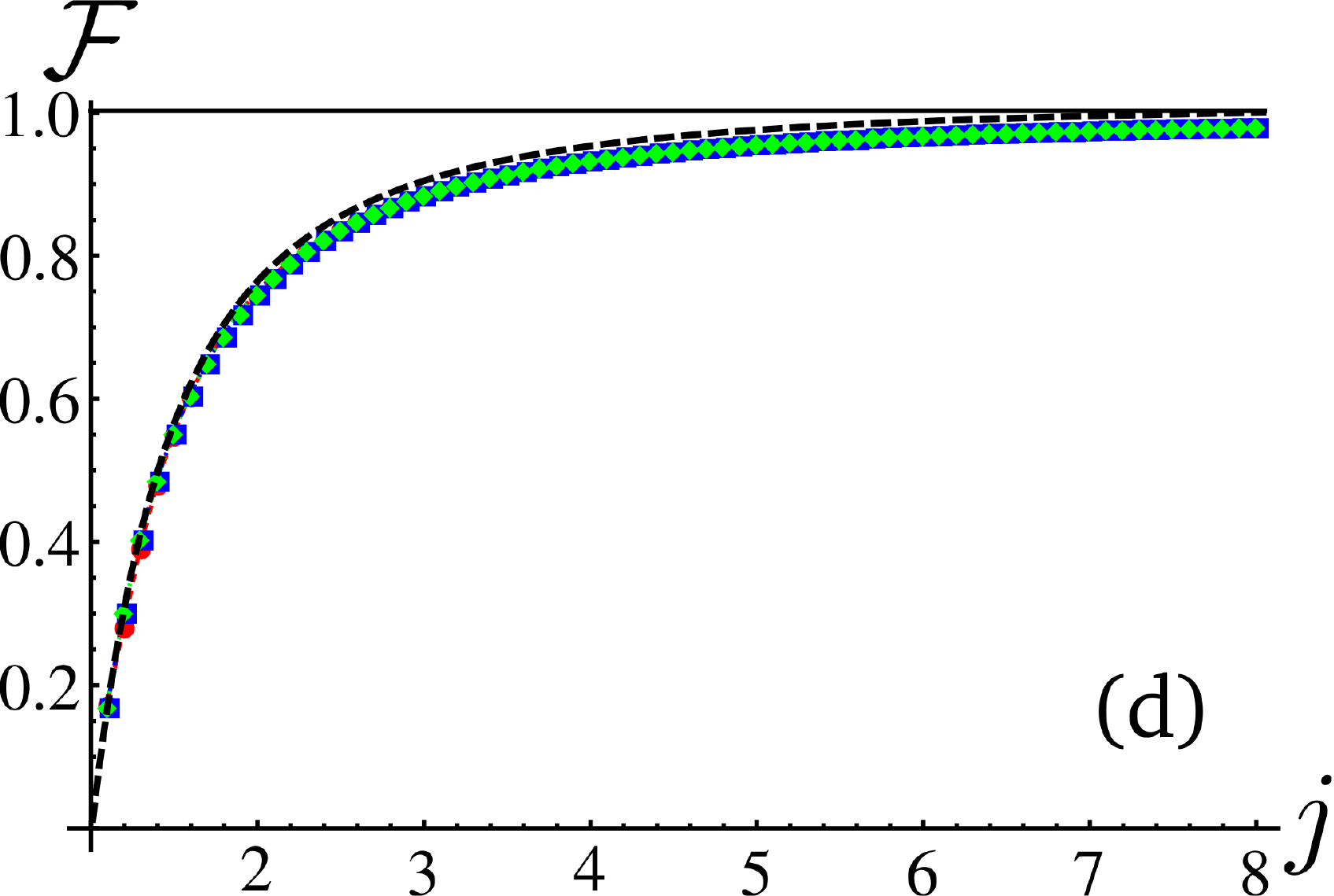}
\caption{(Color online) Fidelity $\mathcal F\!\left(\ket{\Sigma}\!,\!\rho\right) $ between the reduced state $\rho$ of our
model and the pure state $\ket{\Sigma}$ of a linear chain with the same number of spins by varying the coupling strength $j>1$ at different values of the magnetic field $h=0,\frac{1}{2},\frac{1}{\sqrt{2}}, 1$ (panels (a), (b), (c), and (d) respectively). We have taken $M=10,100,1000$ in all panels. The dashed line shows the behavior of the function $1-1/j^2$, which matches the thermodynamic limit of the state fidelity at large magnetic fields.}
  \label{F.fidelitycomposta}
\end{figure}

In Fig.~\ref{F.fidelitycomposta} the fidelity is shown as a
function of $j>1$, for different values of $h$ and $M$. As a
perturbative analysis suggests, for  $j\gg1$ the ground state of
our model tends to the factorized state
$\ket{\Psi^+}_{\pm\frac{1}{2}}\otimes\ket{\omega}_{-\frac{3}{2},\ldots\frac{3}{2}}$,
where $\ket{\Psi^+}_{\pm\frac{1}{2}}$ is a Bell state of the spins
across the BI, while
$\ket{\omega}_{-\frac{3}{2},\ldots\frac{3}{2}}$ is a pure state of
the rest of the system. Fig.~\ref{F.fidelitycomposta} shows that,
almost independently of the magnetic field value, the mixed state
of the reduced system is almost indistinguishable from the target
state for relatively small values of the impurity strength. As far
as finite-size effects are involved, we note that the shorter the
ring, the lower the value of $j$ needed for cutting it, although
differences decrease with increasing $j$ and $h$ [see
Figs.~\ref{F.fidelitycomposta}~(a)-(c)]. On the other hand, for
$h\geq1$ finite-size effects are almost absent but for $j\lesssim2$
[cf. Fig.~\ref{F.fidelitycomposta}~(d)]. This can be easily
explained by noticing that the target state is fully polarized,
$\ket{\Sigma}=\ket{0}^{\otimes 2(M-1)}$, while the ground state of
the ring is $\ket{\Omega}{=}\hat\zeta_-^\dagger\ket{0}$. When the
localization length $q^{-1}$ is less than the length of the ring
2$M$, by taking into account Eq.~\eqref{E.locstates} we get that
the spins located at a distance $d>q^{-1}$ are, for all practical
purposes, in state $\ket{0}$. As a consequence, considering longer
chains will not affect substantially  the value of the fidelity
due to the presence in the ground state of our model of only a
single localized mode. This is at variance with the case $h<1$
where the extended (distorted) eigenstates given by
Eq.~\ref{E.modes}, spread all over the chain. Therefore, for
$h\geq1$ the length of the ring does not play a significant role.
Moreover, the analytical expression for the fidelity in the
thermodynamic limit reads
$\mathcal{F}=1\,{-}\,e^{-2q}=1\,{-}\,1/j^2$. It is worth noticing
that, for all practical purposes, the thermodynamic limit is
already reached when the length of the chain exceeds the
localization length $q^{-1}=1/(\ln{j})$. Finally, for arbitrarily
large values of $h$, the target state does not change because the
XX-Heisenberg model enters the saturated phase. In addition, as
the localized mode is independent of $h$, the ground state of our
model is invariant for $h\geq 1$. This yields the very same
behavior of the fidelity for $h>1$ as that reported in
Fig.~\ref{F.fidelitycomposta}~(d).

\section{Conclusions}\label{S.conclusions}

We have shown that, by means of a BI, it is possible to turn a spin
chain with PBC into an Open Boundary one. The XX-impurity model has been solved
analytically in the thermodynamical limit and two-points magnetic
correlations functions, as well as CC and QD, have been shown to
decay to zero for spin residing across the BI already for a
relatively modest value of the impurity strength. The analogous
figures of merit for pairs of spins residing on the same side of the
BI take values approaching those of a chain with OBC. For finite, yet
arbitrarily large, spin chains, the fidelity between the ground state
of a chain including all the spins but those coupled by the BI and an
open chain of the same size, has been adopted in order to confirm the
validity of the approach discussed here. It follows that impurity
bonds can be used in otherwise translation invariant systems as a
means to achieve an effective cutting of the spin chain at the
desired point. The full analytical treatment provided here allows for
an exact quantification of the cutting quality.

This result shows the possibility, via impurity bonds, to break-up physical systems with a ring topology or to cut long chains in smaller ones by different specific techniques depending on the actual physical implementation, such as chemical doping in molecular spin arrays~\cite{Timco2009,TroianiBCLA2010}, site-dependent modulation of the trapping laser in cold atoms/ions systems~\cite{Lewensteinetal07} or spatial displacement of an optical cavity in an array~\cite{GiampaoloI2010}. This could be exploited in order to make some systems more useful for quantum-state transfer, where often a necessary requisite consists in an addressable head and tail as well as in the finiteness of the quantum data bus. Finally, tuning the values of the impurity strength within $j\in\left[0,10\right]$ is sufficient to investigate the emergence of edge effects, such as total or partial wavefunction backscattering which, by choosing an appropriate uniform magnetic field, spatially modulate the spin correlations functions.

\acknowledgments
{TJGA is supported by the European Commission, the
European Social Fund and the Region Calabria through the program POR
Calabria FSE 2007-2013 - Asse IV Capitale Umano-Obiettivo Operativo
M2.
LB is supported by the ERC grant PACOMANEDIA.
MP thanks the Alexander von Humboldt Stiftung, the UK EPSRC for a Career Acceleration Fellowship and a grant under the ``New Directions for EPSRC Research Leaders" initiative (EP/G004759/1), and the John Templeton Foundation (grant ID 43467).}

\appendix
\section{Diagonalization of the Hamiltonian}\label{a.diag}
We introduce the $2M$ discretized {wavevectors}
$k\equiv{\pi{\ell}}/{M}$ $(\ell=-M{+}1,...,M)$ and the fermionic
operators
\begin{equation}
\hat c_k{=}\frac{1}{\sqrt{2M}}\sum_n e^{i n k}\hat c_n,
\end{equation}
corresponding to excitations of energy $E_k=2(\cos{k}-h)$. States
with one fermionic excitation of (unperturbed) energy $E_k$ are
$\ket{k}{=}\hat c^\dag_k\ket{0}$, where $\ket{0}$ is the fermionic
vacuum state. In this appendix, the analysis is restricted to the
single particle sector of the full Fock state, spanned by these
states. Due to the non-interacting form of the Hamiltonian, the
diagonalization performed in this one-particle sector allows to
straightforwardly obtain the full many-fermion energy
eigen-states. The Green operator of the unperturbed (one-particle)
Hamiltonian is thus defined as
\begin{equation}
\label{G0}
 \hat G_0(z)\,{=}\,\frac{1}{z-\hat{\cal H}_0}
 \,{=}\,\sum_k\frac{1}{z-E_k}\ket{k}\bra{k}~~(z\in\mathbb C).
\end{equation}
In the thermodynamic limit ($M\rightarrow\infty$) the summation is
changed into an integral and the discrete energies $E_k$ become a
continuous energy band. The matrix elements of the Green operator in
the lattice position space read
\begin{equation}
\begin{aligned}
 G_{0}(n,m;z)&=\frac{\left(-x+\sqrt{x^2-1}\right)^{|n-m|}}{2\sqrt{x^2-1}}
  ~~\mathrm{for}~~z\notin I_b,
\\
 G^{\pm}_{0}(n,m;z)&{=}\frac{\left(-x\pm i \sqrt{1-x^2}\right)^{|n-m|}}
 {\pm 2 i \sqrt{1-x^2}}~~~\mathrm{for}~z\in I_b,
\end{aligned}
\end{equation}
where $x=z/2+h$, while ${I_b}=[-2h{-}2,-2h+2]$ is the unperturbed
energy band. The Green operator $\hat G(z)$ associated with the
(one-particle restriction of) the Hamiltonian in
Eq.~\eqref{e.hamiltonf} can be now obtained by the relation $\hat
G(z){=}\hat G_0(z){+}\hat G_0(z)\hat T(z)\hat G_0(z)$, where the
matrix $\hat T(z)\,{=}\,\left(\sum^\infty_{l{=}0}[\hat{\cal
H}_{\rm{I}} \hat G_0(z)]^l\right) \hat{\cal H}_{\rm{I}}$ can be
analytically summed up to all terms. Finally, the knowledge of
$\hat G(z)$ allows us to obtain the whole (single-particle)
spectrum of the Hamiltonian, which consists of the above-mentioned
energy band and a pair of out-of-band discrete energy eigenstates,
which are simple poles of $\hat G(z)$ appearing only for $j>1$. In
order to obtain the corresponding eigenstates, we use the relation
$\ket{\Psi_E}{=}\left[\openone{+}\hat G^+_0(E)\hat
T^+(E)\right]\ket{k}$ for the continuous in-band states, which
describe distorted spin waves of the system that are built from
the unperturbed ones by including the corrections due to the
scattering from the defect and described by the retarded Green
operator $\hat G^+_0(z)$ and the $\hat T(z)$ operator. The
Schr\"{o}dinger equation of the full problem is then solved by
using an appropriate ansatz for the two discrete out-of-band
energy eigenstates~\cite{Pury1991}.

\section{Ring cut fidelity}\label{a.fid}
The ring cut fidelity is easily obtained from the explicit expression of
the ground states \eqref{e.diracsea}. In order to elucidate the main steps
of this derivation let us consider two sets of Fermi operators
$\hat\chi_k = \sum_{n} V_{kn} \hat c_n$,
$\hat\xi_k = \sum_{n} U_{kn} \hat c_n$. The fidelity between two
{\it Dirac seas} follows then from Wick's theorem
\begin{align}
  \bra 0 \prod_{k=1}^{K_F} \chi_k \prod_{k'=1}^{K_{F'}}
  \xi_{k'}^\dagger\ket 0 &=
  \begin{cases}
    0 & ~\text{if}~K_F\neq K_{F'}~,\\
    \det G & ~\text{if}~K_F=K_{F'}~,
  \end{cases}
  \label{e.fidasd}
  \\
  G_{kk'} = \bra 0 \chi_k\xi_{k'}^\dagger\ket 0 &= \sum_n V_{kn} U_{k'n}^*~.
\end{align}
The ring cutting fidelity then reads
\begin{align}
F=&\bra\Sigma\Tr_{n{=}{\pm\frac{1}{2}}}\left[\ket{\Omega}\!\!\bra{\Omega}\right] \ket\Sigma =
\\\nonumber=&|\Sprod{\tilde\Sigma}{\Omega}|^2 +
|\SprodO{\tilde\Sigma}{c_{-\frac{1}{2}}}{\Omega}|^2 +\\\nonumber&
|\SprodO{\tilde\Sigma}{c_{+\frac{1}{2}}}{\Omega}|^2 +
|\SprodO{\tilde\Sigma}{c_{-\frac{1}{2}}c_{+\frac{1}{2}}}{\Omega}|^2~,
  \label{e.fidelity}
\end{align}
where $|{\tilde\Sigma}\rangle$ refers to the state $\ket{\Sigma}$ extended to
the larger Fock space of $2M$ Fermions. Each term of the above sum is then
evaluated from \eqref{e.fidasd} with a suitable choice of the matrices
$V$ and $U$.

\end{document}